\pdfoutput=1
\documentclass[aps,prl,twocolumn,superscriptaddress]{revtex4}

\usepackage{graphicx}
\usepackage{boxedminipage}

\begin{document}

\title{Proposal of a one-dimensional electron gas in the steps at the LaAlO$_3$-SrTiO$_3$ interface}

\author{N. C. Bristowe}
\affiliation{Theory of Condensed Matter,
             Cavendish Laboratory, University of Cambridge, 
             Cambridge CB3 0HE, UK}
\affiliation{Department of Earth Sciences, University of Cambridge, 
             Downing Street, Cambridge CB2 3EQ, UK}
\author{T. Fix}
\affiliation{Department of Materials Science, University of Cambridge, 
             Pembroke Street, Cambridge CB2 3ZQ, UK}
\author{M. G. Blamire}
\affiliation{Department of Materials Science, University of Cambridge, 
             Pembroke Street, Cambridge CB2 3ZQ, UK}
\author{P. B. Littlewood}
\affiliation{Theory of Condensed Matter,
             Cavendish Laboratory, University of Cambridge, 
             Cambridge CB3 0HE, UK}         
\affiliation{Physical Sciences and Engineering,
             Argonne National Laboratory, 
             Argonne, Illinois 60439, USA}    
\author{Emilio Artacho}
\affiliation{Theory of Condensed Matter,
             Cavendish Laboratory, University of Cambridge, 
             Cambridge CB3 0HE, UK}
\affiliation{CIC Nanogune, and Donostia International Physics Center DIPC,
		Tolosa Hiribidea 76, 20018 San Sebastian, Spain}      
\affiliation{Basque Foundation for Science Ikerbasque, 48011 Bilbao, Spain}
\affiliation{Department of Earth Sciences, University of Cambridge, 
             Downing Street, Cambridge CB2 3EQ, UK}

\date{\today}
\begin{abstract}
The two-dimensional electron gas (2DEG) at the interface between LaAlO$_3$ (LAO) 
and SrTiO$_3$ (STO) has become one of the most fascinating and highly-debated oxide systems of recent times. 
Here we propose that a one-dimensional electron gas (1DEG) can be engineered at the step edges 
of the LAO/STO interface.
%In addition to the 1DEG, a variety of electronic phases are predicted. 
These predictions are supported by first principles calculations and electrostatic modeling
which elucidate the origin of the 1DEG as an electronic reconstruction 
to compensate a net surface charge
%of $\pm$$e$/2$S$
in the step edge.
%, in addition to the terrace.
The results suggest a novel route to increasing the functional density in these electronic interfaces.
\end{abstract}
\maketitle
 
% \section{Outline}

%{\bf Introduction}: LAO/STO switchable 2DEG - potential applications in technology.
%Literature - electronic reconstruction since LAO has a polarization of half a quantum (or equivalent charged planes argument).
%Epitaxial growth of LAO-STO, always produces steps (as seen by AFM)
%Any unusual electronic behavior in steps already in literature?

%{\bf Model}: Figure 1. Polarization picture: LAO has a polarization of $\pm P_0$ (half a quantum) along x, y and z. 
%Charged planes picture: The terrace and step edges have "chemical" charge densities, $\pm\sigma_c$. 
%Both pictures produce a net interface charge density, $\tilde{P}$ or $\tilde{\sigma}_c$ of $P_0($cos$\theta \pm $sin$\theta)$,
%with $\theta$ the miscut angle. The $\pm$ depends on the step termination: + if the step and terrace have the same terminations,
%and - if they have different terminations. Various phases are expected depending on the relative terminations, and LAO thickness which are %discussed below. 

%The study of novel electronic phases at oxide interfaces has rocketed since 
%the discovery of 
The two-dimensional electron gas (2DEG) at the interface
between two band insulators,  LaAlO$_3$ (LAO) and SrTiO$_3$ (STO),
has become the prototypical system for the study of novel electronic phases at oxide interfaces. 
Since the original discovery~\cite{Ohtomo2004} it has been shown that
at low temperatures
the 2DEG may become magnetic~\cite{Brinkman2007}, superconducting~\cite{Reyren2007}, 
or both coexisting in one phase~\cite{Bert2011,Li2011}.
%the 2DEG may exist in various phases
%at low temperatures;
%magnetic~\cite{Brinkman2007}, superconducting~\cite{Reyren2007}, 
%or an exotic phase of both magnetic and superconducting
%separated~\cite{Wang2011} or 
%coexisting~\cite{Bert2011,Li2011}.
At room temperature the LAO-STO interface may find applications 
in field-effect devices~\cite{Thiel2006,Cen2008}, sensors~\cite{Xie2011} or nano photo-detectors~\cite{Irvin2010}. 
%Despite the wealth of exciting discoveries in this system, the origin of the 2DEG is still under debate. 

The origin of the 2DEG is still under debate, but
one potential mechanism,
% for the 2DEG origin
called the ``electronic reconstruction"~\cite{Nakagawa2006},
is currently the most popular~\cite{Schlom2011}.
At the heart of this mechanism is the notion that LAO has a non-zero
polarization, $P$~\cite{Stengel2009,Bristowe2011}, which for thin films, translates to 
having polar surfaces/interfaces along certain crystallographic directions~\cite{Vanderbilt1993},
(including (001) - the most common growth direction in practice). 
Polar surfaces are electrostatically unstable but one possible route to charge compensate 
is via a transfer of electrons between the surface of the film and the interface. 
For a LAO film grown on a TiO$_2$ terminated STO (001) substrate, complete compensation
would amount to precisely 0.5 $e/S$ (where $S$ is the area of the (001) plane associated to one formula unit)
transferring from the LAO surface to the interface, creating the 2DEG. 
Incomplete, or no compensation may occur for ultrathin LAO films, where the internal
potential drop is lower than the relevant effective gap (in this case the band gap of STO). 
This is consistent with the observation of a critical LAO thickness, $d_c$, for 2DEG appearance~\cite{Thiel2006}. 
An alternative charge compensating mechanism is via surface oxygen vacancies~\cite{Cen2008}
which is also consistent with the observed $d_c$ (see Ref.~\onlinecite{Bristowe2011a}).

\begin{figure}[b]
\begin{center}
%\begin{boxedminipage}{0.47\textwidth}
\includegraphics[width=0.45\textwidth]{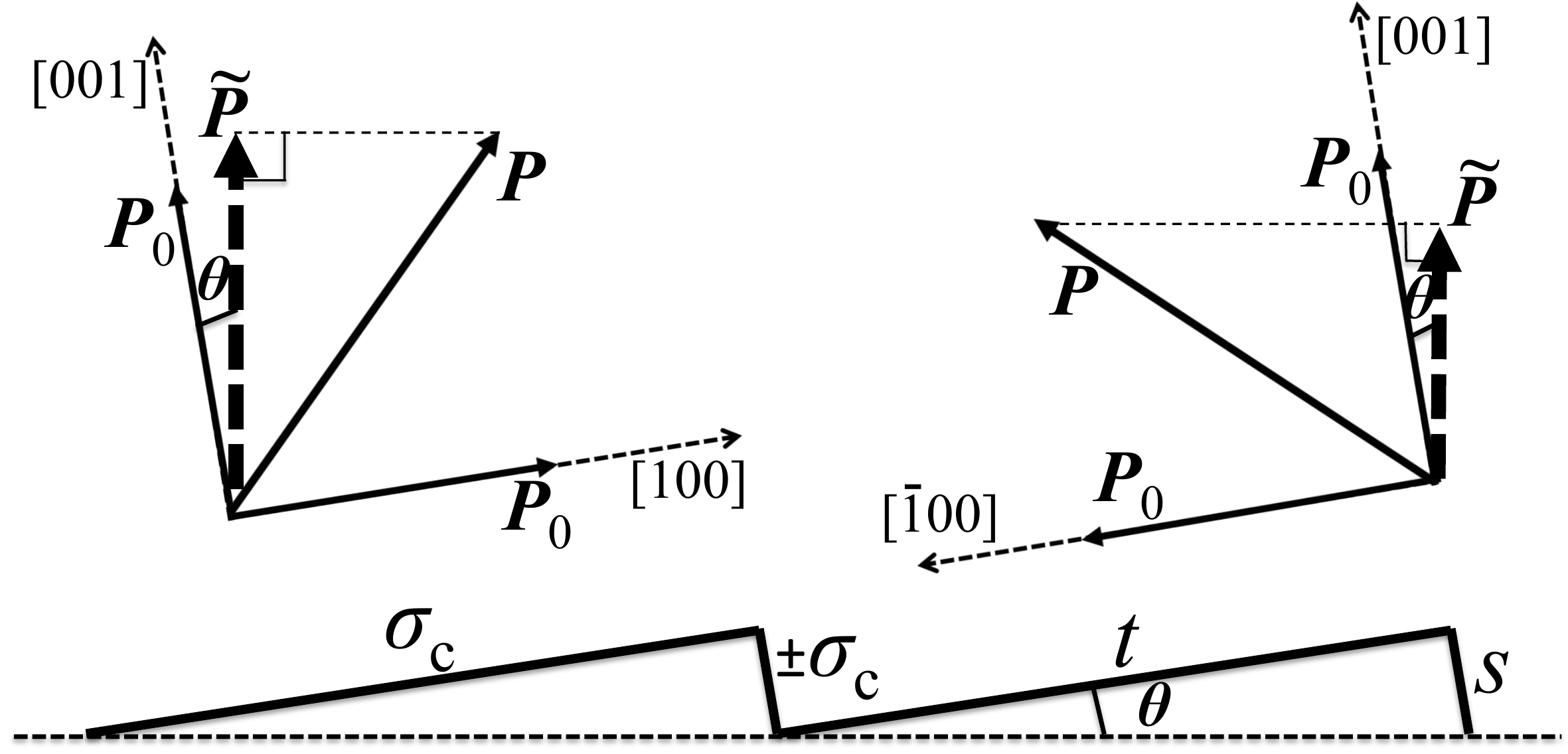}
%\end{boxedminipage} 
%\includegraphics[
%width=0.45\textwidth,viewport= 50 5 800 620,clip]{3b}
\caption{\label{TRAP}{
Schematic diagram illustrating the effect of steps on the net charge of the interface.
Top: Polarization argument considering LAO
has a polarization, $P_0$ along the [100], [010] and [001] directions,
with $\tilde{P}$ the net of the stepped interface.
Bottom: an equivalent picture with charge densities $\pm\sigma_c$ on the
terrace and step edge. 
}}
\end{center}
\end{figure}

In this paper we propose that steps at the interface, defects which are observed
in practically all epitaxial films (due to a small miscut  of the substrate by angle $\theta$), 
can be utilized to produce one-dimensional electron gases (1DEG)
residing at the step edge. 
We use a simple model to argue that a step of one unit cell in height alters
the polarity of the interface and surface of LAO, since the step edge itself has a net charge
of $e/2S$ (i.e. a charge line density of $s e / 2 S$, with $s$ the 
step height as in Fig. 1).
%equal magnitude to the terrace, $e/2S$.
An equivalent, and more rigorous, picture takes the component perpendicular to the stepped surface
of the formal polarization of LAO (Fig. 1), which is precisely (1,1,1) $e/2S$ as shown later in the text.
%(i.e. $\sqrt{3}$ $e/2S$ along the [111] direction, as shown later in the text)
%, modulo a quantum of polarization (Fig. 1).
The electrostatics of the interface are studied with a simple capacitor plate model,
supported with density functional theory (DFT) calculations, that shows an alteration of the 
interface carrier density with miscut angle.
A 1DEG is predicted for small miscut angles when the LAO thickness is just below $d_c(\theta=0)$,
the critical thickness for metallicity in the zero miscut system.
% (which we call the pristine system from now).

\begin{figure}[t]
\begin{center}
\includegraphics[
totalheight=0.32\textheight,
width=0.17\textwidth,viewport= 30 -70 560 920,clip]{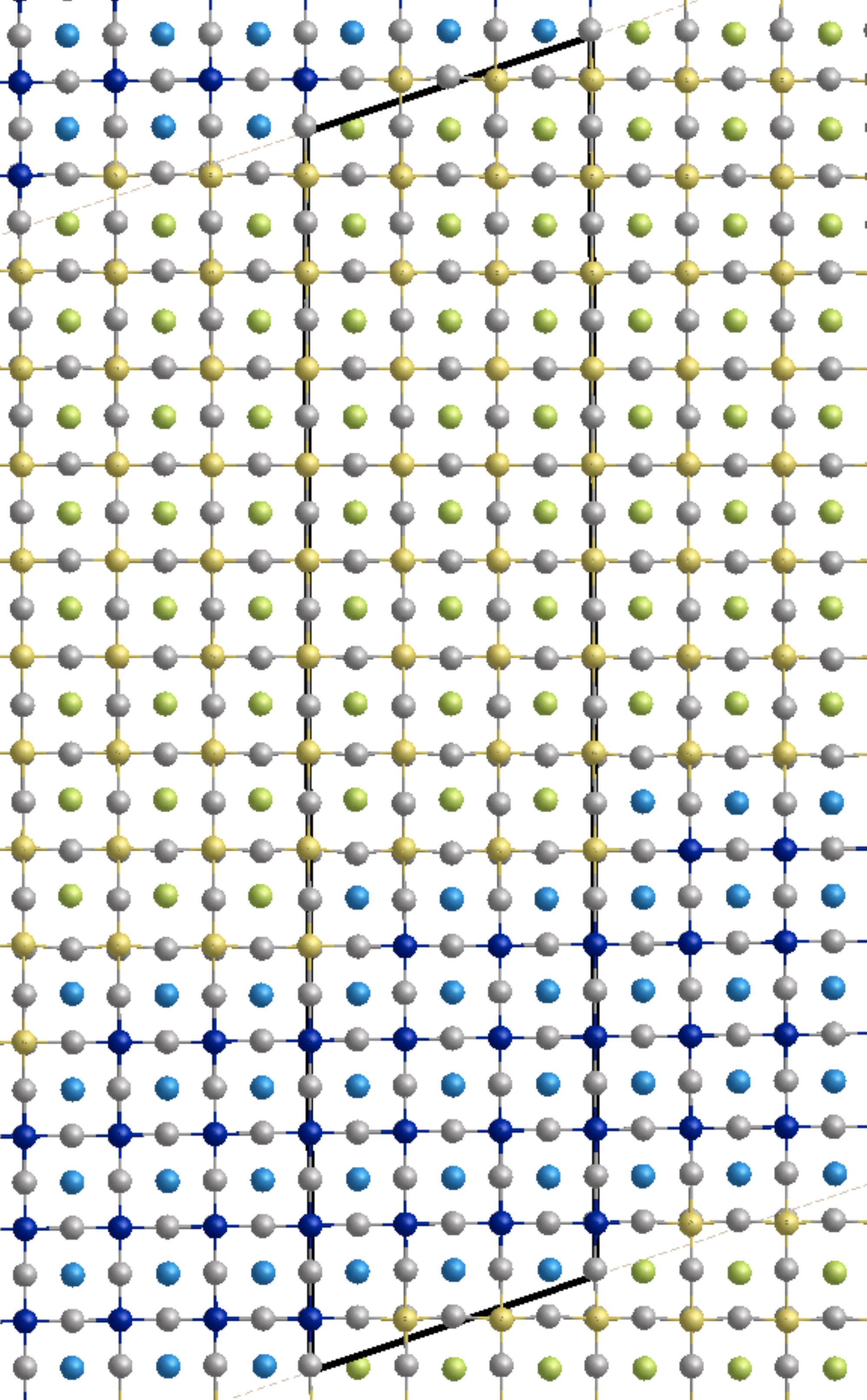}
\includegraphics[
%totalheight=0.25\textheight,
width=0.30\textwidth,viewport= 40 0 610 790,clip]{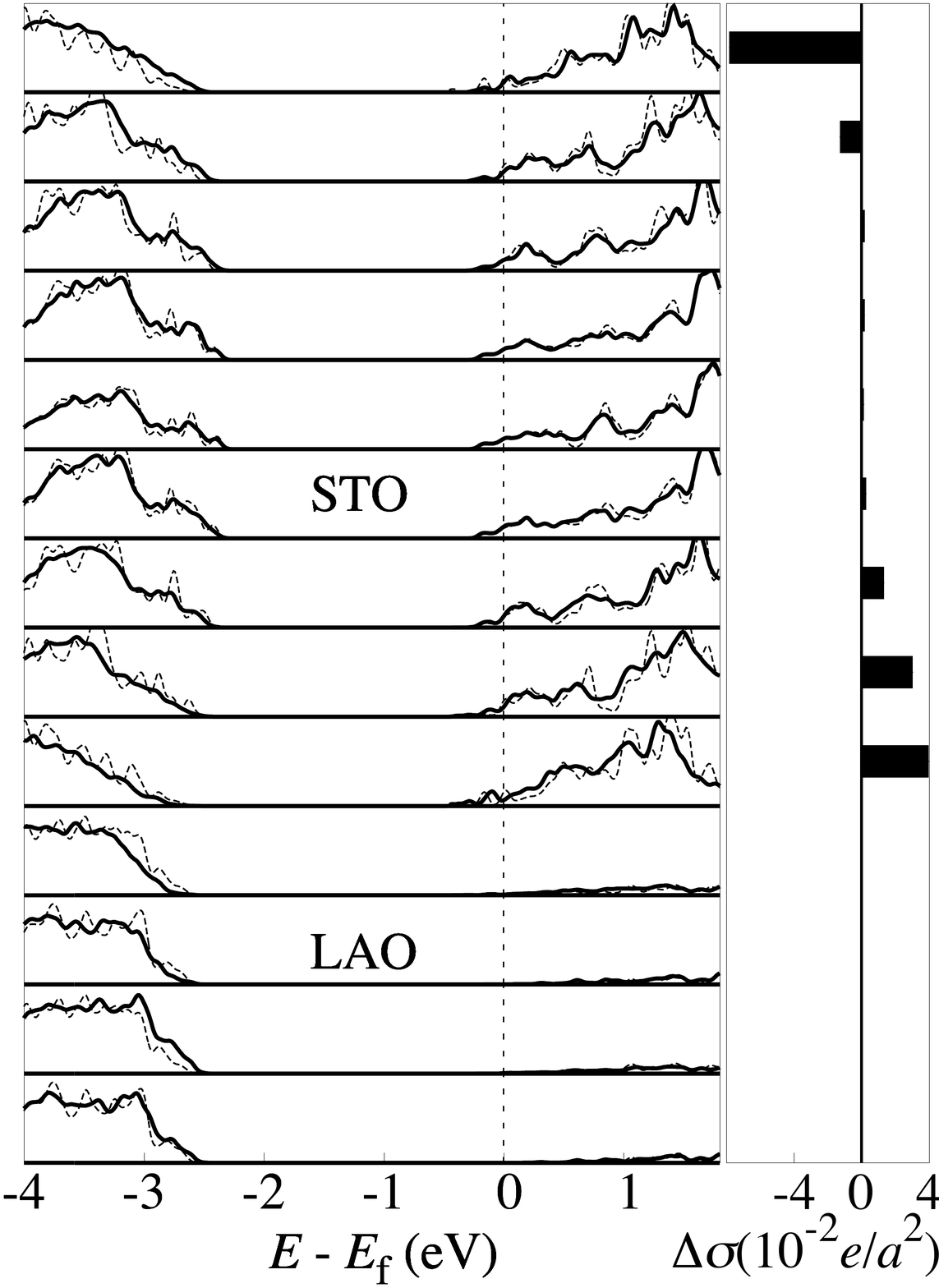}
\caption{\label{TRAP}{
(Color online) Left: The DFT simulation cell of the LAO/STO superlattice. 
The two interface terrace terminations are both TiO$_2$-LaO ($n$-type) whereas the 
interface step termination is $n$-type along the bottom interface and $p$-type
(SrO-AlO$_2$) along the top interface.
(grey: oxygen, dark blue: Al, light blue: La, yellow: Ti, green: Sr)
Center: (001) Layered density of states near the Fermi level of the stepped superlattice 
(solid lines) and pristine superlattice (dashed line).
Right: The change in the layer density of free electrons 
on introduction of steps in the superlattice. 
%from the stepped
%to the pristine superlattice.
}}
\end{center}
\end{figure}

We begin with a formal consideration of the polarity of stepped LAO surfaces. 
We define the miscut angle as tan$\theta=s/t$, where $s$ and $t$ are the step 
and terrace distances respectively (see Fig. 1). 
The steps can be intentionally miscut along a certain direction~\cite{Brinks2011,Fix2011}.
We consider the [100] case here,
but the arguments can be trivially applied to any miscut orientation. 
Atomic force microscopy (AFM) measurements of epitaxial LAO-STO suggest steps of 1 unit cell in height~\cite{Brinks2011,Fix2011}.
Neighboring terraces of the surface are then of the same termination (BO$_2$ here),
but the step edge may be terminated either AO or BO$_2$ (non-polar vicinal LAO surfaces have been considered in Ref.~\onlinecite{Stengel2011c}).
Perhaps the conceptually simplest picture is to take what has been called ``chemical"~\cite{Bristowe2009} (or
``compositional"~\cite{Murray2009a}) charges, $\sigma_c=0.5 e/S$, along the terrace and step edges (Fig. 1 bottom). 
%We will later show this picture produces the same result with the modern theory of polarization. 
The total net charge of the surface,  $\tilde{\sigma}_c$, is then $\sigma_c($cos$\theta \pm $sin$\theta)$.
The $\pm$ depends on the step termination: addition if the step and terrace have the same terminations,
and subtraction if they have opposite terminations. 
%(the difference being exactly a quantum of polarization). 
This result can be confirmed within the framework of the modern theory of polarization~\cite{Kingsmith1993}. 
Taking the formal polarization, $P$, defined by~\cite{Resta1994} the position of the ion cores and mapping the electronic 
wavefunction on to Wannier centers~\cite{Marzari1997}, one finds for cubic LAO  $P=\sqrt{3}$ $e/2S$ along the [111] direction~\cite{Stengel2011c}.
The component of this polarization normal to the stepped surface (Fig. 1 top), $\tilde{P}$,
is the net surface charge density and precisely the same as the above ``chemical" charge.
For tetragonal LAO (as in the case of epitaxial thin films), a trivial correction may be applied to 
the above through a redefinition of $S$.

To support this picture, density functional theory (DFT) calculations were performed on two model LAO-STO 
systems: $n$-$n$ superlattices, and $n$-$p$ films.
$n$ and $p$ denote the termination of LAO terraces (n - LaO, p - AlO$_2$).  
Note that in both cases, the steps are created by simply altering one of the supercell vectors - 
there is no change in stoichiometry through the introduction of steps.
The DFT calculations were performed using the Wu-Cohen exchange correlation functional~\cite{Wu2006a}
as implemented in the  {\sc Siesta} code~\cite{Ordejon1996,Soler2002}. 
Details of the pseudopotentials and atomic-orbitals basis are given in Ref.~\onlinecite{Bristowe2009}. 
We begin with $n$-$n$ superlattices (Fig. 2).

Fig. 2 shows the effect of steps on the electronic structure of the $n$-$n$ superlattice. 
The superlattice was constructed from 4.5 unit cells of LAO and 8.5 unit cells of STO
stacked along the [001] direction.
Steps were introduced by taking a supercell 
%(with $t$ repeat units along the [100] direction),
and altering the {\bf a} supercell vector to $t${\bf a}+$s${\bf c}
% to [t0s]
%by s unit cells along the [001] direction 
(see Fig. 2 left). 
This produces $n$-type steps along one interface and $p$-type steps along the other while maintaining exactly the same stoichiometry.
Fig. 2 shows the case of $t$=3 and $s$=1. 
In the system without steps (which we call the pristine system) the Ti 3$d$ conduction electrons distribute evenly between the two $n$-type terraces
- 0.5 $e/S$ at each interface with a decay through the STO slab (Fig. 2 center),
consistent with previous DFT calculations~\cite{Janicka2009,Popovic2008}.
The inclusion of steps
is found to transfer electrons from one interface to the other (Fig. 2 right), 
accumulating near the $n$-type steps, and depleting near the $p$-type steps,
with respect to the pristine system.
The total density of electrons transferred is approximately 0.1$e/S$ (20\% of $\sigma_c$).
This is consistent with the number required to screen the bound charge density
of each interface, including the step charges, as predicted by the model. 
This model system clearly shows a large effect of steps on the interface free electron density and distribution. 
Note that there has been no change in stoichiometry nor is there any other mechanism suggesting an additional charge transfer.

\begin{figure}[t]
\begin{center}
\includegraphics[
width=0.45\textwidth,viewport= 90 40 770 540 ,clip]{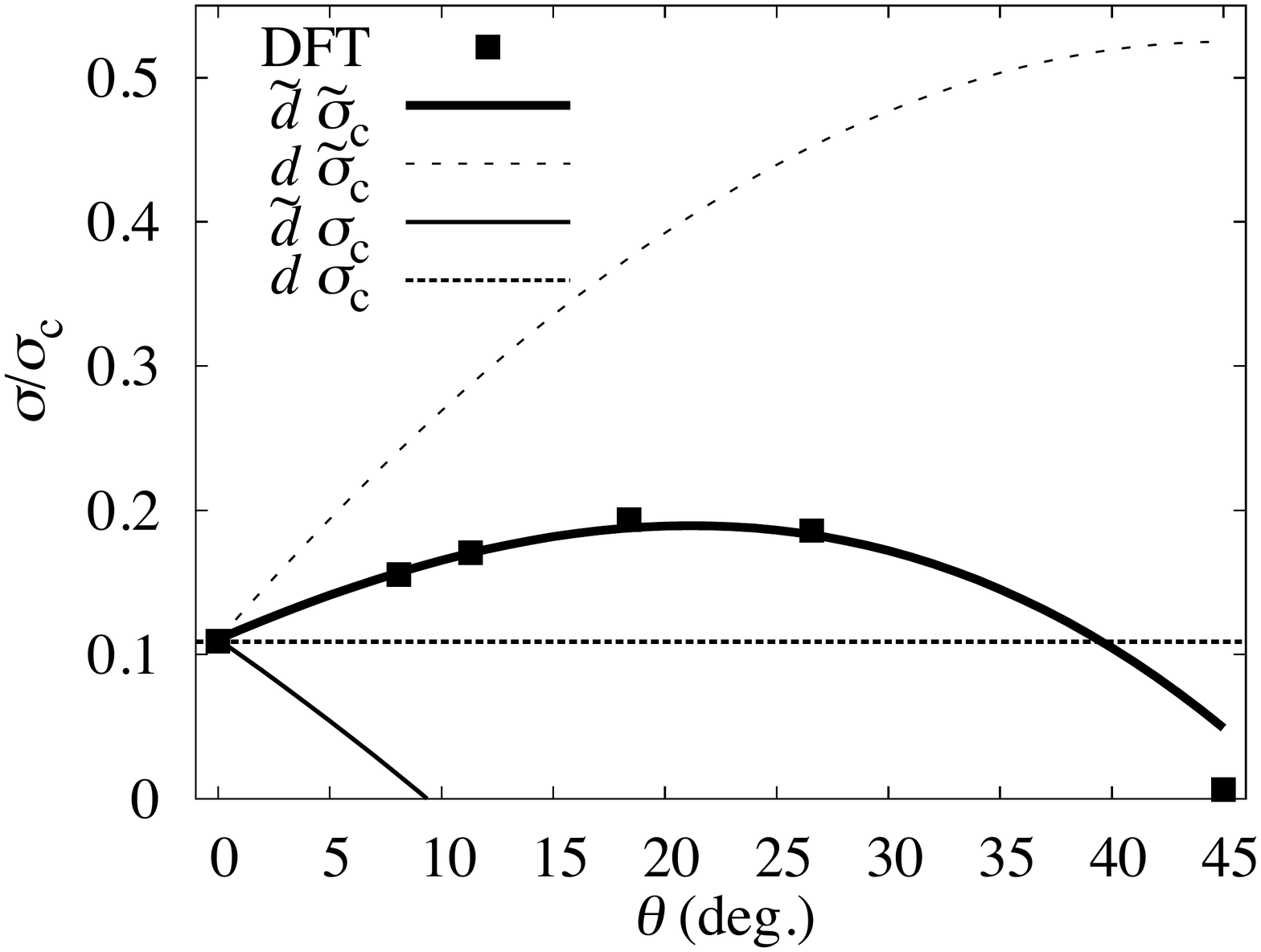}
%\begin{boxedminipage}{0.4\textwidth}
%\end{boxedminipage} 
\includegraphics[
width=0.4\textwidth,viewport=  20 10 700 470,clip]{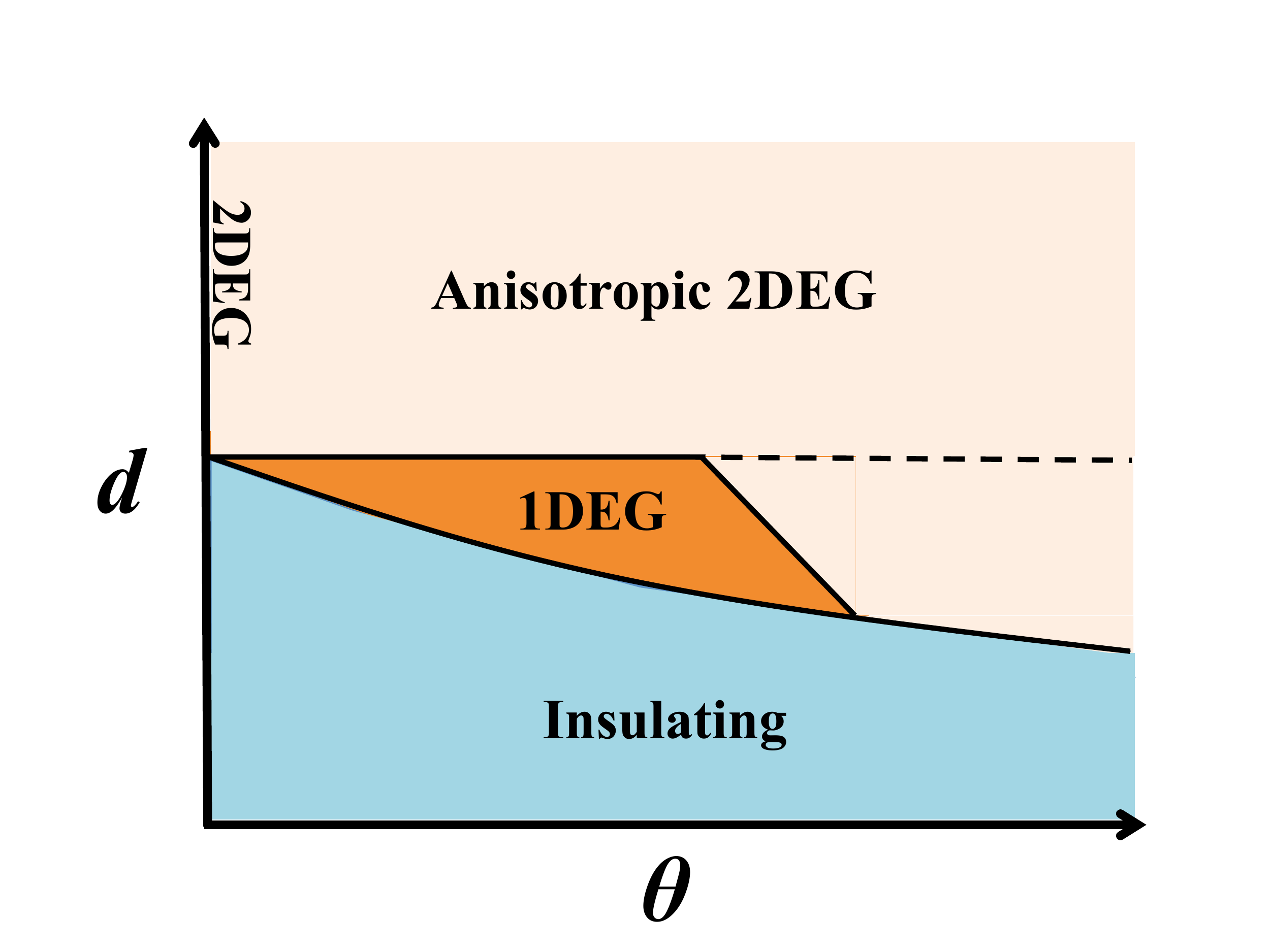}
\includegraphics[
width=0.4\textwidth,viewport= 0 0 1000 600,clip]{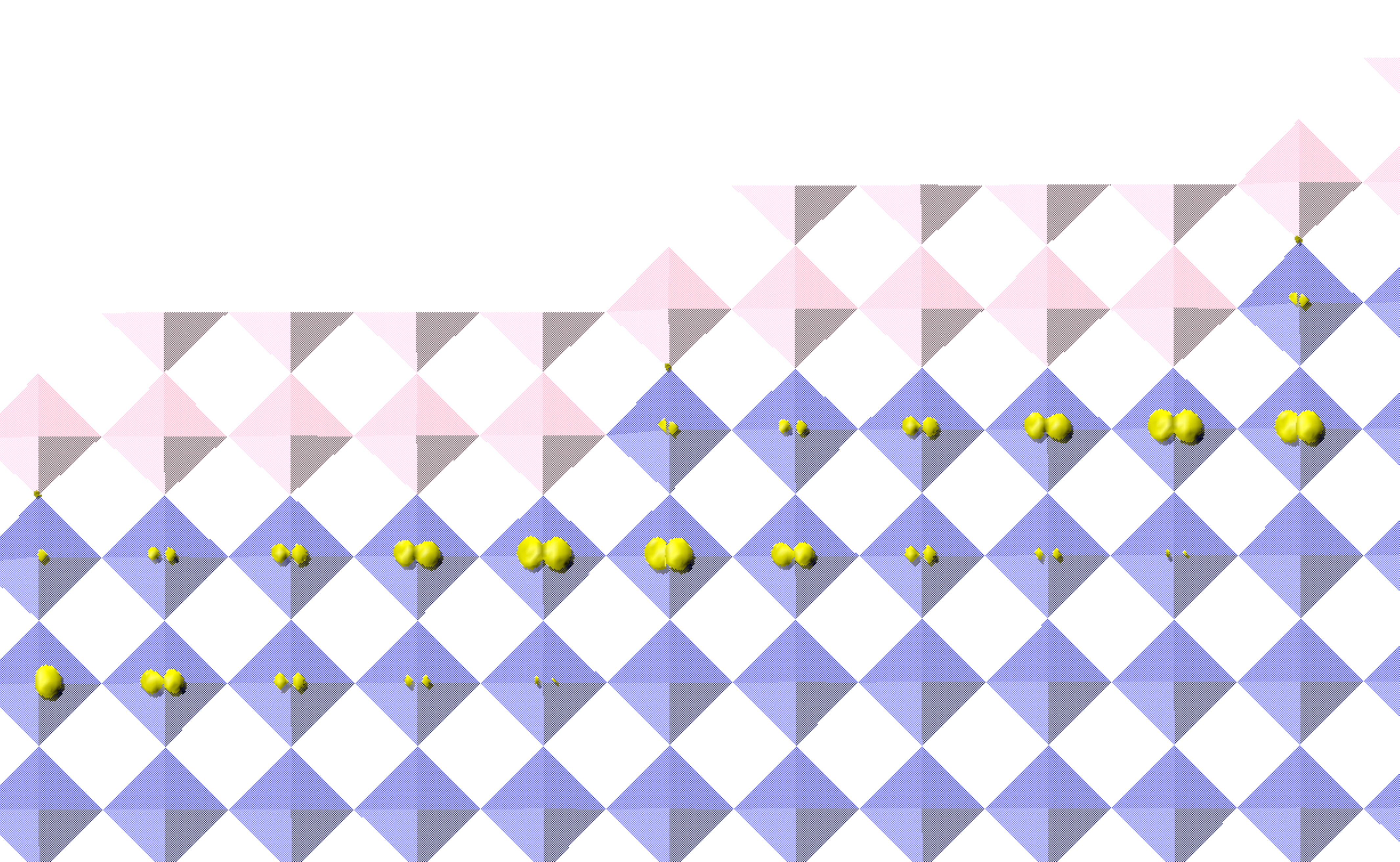}
\caption{\label{TRAP}{
(Color online) 
Top: Interface free electron density, $\sigma$, as a function of the miscut angle, $\theta$,
for a LAO(3 unit cells)/STO film geometry. 
Points represent densities calculated from DFT, and lines are electrostatic modeling under various 
assumptions (see text).
Middle: Predicted schematic electronic phase diagram at the $n$-type (step and terrace) terminated
LAO/STO interface as a function of the miscut angle, $\theta$, and the LAO thickness, $d$. 
Bottom: An isosurface of the Ti 3$d$ electron density (yellow clouds) at an $n$-type stepped
LAO/STO interface. Blue and pink cages represent oxygen octahedra around Ti and Al atoms 
respectively.
%and pink cages represent oxygen octahedra around Al atoms.
}}
\end{center}
\end{figure}

We now consider a system of more relevance for most experimentally grown systems in reality - $n$-$p$ LAO films
on STO substrates. 
The system consists of a slab geometry of $d$ LAO unit cells stacked on 5 unit cells of STO along the [001] direction
including a 15 \AA\ vacuum layer. 
A dipole correction was included in the vacuum layer to simulate open-circuit boundary conditions, preventing
unphysical macroscopic electric fields due to the asymmetry of the slab and the periodic boundary conditions. 
The introduction of steps was the same as in the above superlattice case,
and chosen such that $n$-type steps reside in the $n$-type terrace, and $p$-type steps reside in the $p$-type terrace. 
We begin with the case of $d=3$ unit cells, which was found to be beyond the pristine critical thickness within DFT for the onset of 
metallicity, and then alter the miscut angle, $\theta$, through varying $t$ ($s$ is again fixed at 1 unit cell to resemble 
that seen in experiments).
The total interface free electron density, $\sigma$, is calculated from the projected density of states, which allows one
to easily separate free and bound charge due to the existence of a gap. 
The results are depicted as points in Fig. 3 top, which shows a clear variation of free charge density with miscut angle,
going through a maximum (nearly 200\% of $\sigma(\theta=0)$) at intermediate angles.
Three curves are compared in Fig. 3 top, each with various model approximations, which we discuss next. 

To model $\sigma$ as a function of $\theta$ we use ($i$) Gauss' Law, $\tilde{\sigma}_c-\sigma=\epsilon${\Large $\varepsilon$}, 
where $\epsilon$ is the LAO dielectric constant and {\Large $\varepsilon$} the LAO internal electric field,  and ($ii$)
voltage drop pinning, {\Large $\varepsilon$}$=\Delta/\tilde{d}$~\cite{Bristowe2009} ($\Delta$ is the STO band gap), obtaining
\begin{equation}
\sigma =  \tilde{\sigma}_c - \frac{\epsilon\Delta}{\tilde{d}}
\end{equation} 
with $\tilde{\sigma}_c = \sigma_c($cos$\theta+$sin$\theta)$ the net charge of the stepped interface 
as discussed earlier, and $\tilde{d}=d(1+C$tan$\theta)/(1+$tan$\theta)$, 
the average shortest distance between the stepped plates ($C$ is a geometrical constant depending only on the ratio $s/d$). 
Eq. 1 produces the bold line curve in Fig. 3 top, which is in good agreement with the DFT calculations. 
Importantly, both the effect of the steps on the net polarization, $\tilde{\sigma}_c$, and on the effective
LAO thickness, $\tilde{d}$, is required to agree with the DFT calculations.
Substituting simply the terrace net charge $\sigma_c$ and or the LAO thickness $d$ results in disagreement with DFT (dotted and thin lines Fig. 3 top).

With strong support from the DFT calculations, we explore the model further. 
Firstly the critical thickness for metallicity as a function of miscut angle, $d_c(\theta)$. 
This is determined from Eq. 1 by setting $\sigma$ to zero and rearranging for $d$. 
A schematic of this transition is shown in Fig. 3 middle, for small $\theta$, as the curved line separating the insulating phase. 
The degree of curvature away from horizontal has been exaggerated for clarity.
In fact, using the experimental values for $\epsilon$ and $\Delta$ and  $\sigma_c=0.5 e/S$, it reaches a minimum 
of about 97\% of $d_c(\theta=0)$ at $\theta \approx 17^\circ$. 
Interestingly, this minimum can be altered through epitaxial strain since the step edge charge density relies 
on the out-of-plane lattice parameter. 
Fig. 3 middle shows three other phases above this transition curve - 2DEG, anisotropic 2DEG and 1DEG.
Above the pristine critical thickness (horizontal line) an anisotropic 2DEG is predicted at any non-zero miscut angle. 
This is due to the non-isotropic net charge density of the interface (see later for support from DFT).
Between the pristine critical thickness and $d_c(\theta)$, and for small $\theta$, a 1DEG phase is predicted. 
Introducing steps just below the pristine critical thickness produces carriers in the step edge, to screen
the excess interface bound charge. 
In the limit of very small $\theta$ these are effectively isolated steps.
Considering one single step, the excess chemical charge
is located around the step for the same reason that the chemical 
charge associated to the pristine LAO/STO (001) interface is 
located close to the interface. 
Since the free carrier charge
being transferred is less than the excess chemical charge,
the potential well generated by the latter confines the carriers
within some width around the step in the direction normal to
the step, thereby defining a 1DEG.
%In the limit of very small $\theta$ these are effectively isolated steps,
%and the additional bound charge is located at the step,
%whereby the carriers transferred towards screening it will 
%locate around the step defining the 1DEG. 
A phase transition with increasing miscut angle occurs when
the 1DEGs, with a finite width, begin to overlap with next neighbor steps.
At this point, the 1DEGs are electronically indistinguishable from the anisotropic 2DEG.

In an attempt to observe the 1DEG, DFT calculations were performed on the $n-p$ film system (see above),
with $d$ as 2 unit cells, which is lower than the pristine critical thickness, and t=5, s=1.
Fig. 3 bottom shows an isosurface of the Ti 3$d$ electron density.
The electron density is clearly inhomogeneous, and accumulates near the step edge. 
The density decays from the step edge in to the terrace, but not to zero.
The phase is the anisotropic 2DEG of Fig. 3 middle, below the dotted line. 
By reducing the miscut angle in the simulation, it should be possible to find the 1DEG, however
we are 
%realistically computationally 
limited to not too small miscut angles due to system size. 
%The step and terrace termination at the interface are chosen to be n-type, and the LAO thickness 
%(2 unit cells) is lower than the pristine critical thickness, with the aim of seeing the 1DEG. 
%The electron density accumulates in the step edges, but decays across the terrace, overlapping 
%with neighboring steps - an inhomogeneous 2DEG. 
%These DFT calculations are realistically limited to large miscut angles, but we predict that in 
%experiments by tuning the LAO thickness, miscut angle (and possibly pristine critical thickness 
%through strain or material) it will be possible to produce 1DEGs at the steps. 

Experimental studies of stepped LAO-STO interfaces are few, but we highlight two here.
Anisotropic electrical transport has been observed, but explained by anisotropic carrier mobility, not density~\cite{Brinks2011}.
Interestingly, the strongest anisotropy was observed at lowest miscut angles, in disagreement with their
step scattering model. 
Elsewhere, steps were found to affect the carrier mobility and density of the 2DEG~\cite{Fix2011}.
The carrier density was found to decrease with miscut angle, suggesting the possibility of $p$-type step formation.

%We finally proceed with suggestions for experimental search of the 1DEG, which we hope this study will stimulate. 
We hope this work will stimulate the study of stepped surfaces and the search for 1DEGs.
Likely candidate systems include; ($i$) $n$-type step and terrace terminations,
 ($ii$) a theoretical pristine critical thickness of only just beyond an integer number of LAO unit cells, ($iii$) a maximised curvature of $d_c(\theta)$ and
($iv$) a small miscut angle to disconnect the 1DEGs from successive steps, but not too small since the effect of the step polarization
is then minimized.
In practice this may mean; ($i$) surface chemical treatment to alter the step edge termination, ($ii$) engineering the pristine critical thickness
(through for example strain~\cite{Bristowe2009,Bark2011}, electric field~\cite{Thiel2006,Cen2008}, material~\cite{Perna2010} or atmospheric environment 
if the electronic reconstruction instead occurs via surface chemical redox reactions~\cite{Cen2008,Bristowe2011a}), and ($iii$)
increasing the step bound charge through epitaxial strain, and miscut orientation. 
%(i.e. increasing the substrate lattice parameter).

In conclusion we have studied the effects of steps at the LAO-STO interface on the
electronic structure.
A simple model of polarization and electrostatics is in good agreement with DFT calculations.
The model predicts the existence of a 1DEG in the step edge at the interface.

\begin{acknowledgments}
We acknowledge
%G Catalan, J \'{I}\~{n}iguez, M Bibes, V Garcia, 
%N Mathur, X Moya, J Junquera, C Ocal and S Streiffer for valuable discussions,
the support of EPSRC and computing 
resources of CamGRID and Darwin at Cambridge, the 
Spanish Supercomputer Network and HPC Europa.
PBL acknowledges DOE support under FWP 70069.
\end{acknowledgments}

%\bibliography{STEPS}

\end{document}